\documentclass[runningheads,a4paper]{llncs}
\usepackage{amssymb}
\usepackage{amsmath}
\usepackage{graphicx}
\usepackage{url}
\usepackage{subfigure}
\usepackage[ruled,linesnumbered]{algorithm2e}
\usepackage[english]{babel}
\urldef{\mailsa}\path|{jatin.agarwal, nadeem.moidu}@research.iiit.ac.in|
\urldef{\mailsb}\path|{kkishore, srinathan}@iiit.ac.in|

\newcommand{\keywords}[1]{\par\addvspace\baselineskip
\noindent\keywordname\enspace\ignorespaces#1}
\begin{document}

\mainmatter  % start of an individual contribution

\title{Efficient Range Reporting of Convex Hull}

\titlerunning{Efficient Range Reporting of Convex Hull}

\author{Jatin Agarwal \and Nadeem Moidu \and Kishore Kothapalli \and Kannan Srinathan}

\authorrunning{J Agarwal, N Moidu, K Kothapalli, K Srinathan}

\institute{International Institute of Information Technology, Hyderabad, India\\
\mailsa\\
\mailsb\\}
%\toctitle{On Counting Range Maxima Points in Plane}
%\tocauthor{Anil Kishore Kalavagattu}
%
\maketitle

\begin{abstract}

We consider the problem of reporting convex hull points in an orthogonal range query in two dimensions.
Formally, let $P$ be a set of $n$ points in $\mathbb{R}^{2}$.
A point lies on the convex hull of a  point set  $S$ if it lies on the
boundary of the minimum convex polygon formed by $S$. In this paper, we
are interested in finding the points that lie on the boundary of the
convex hull of the points in $P$ that also fall with in an orthogonal range
$[x_{lt},x_{rt}]\times{}[y_b, y_t]$. 
We propose a $O(n \log^{2} n) $ space data structure that can support reporting points on a convex hull inside
an orthogonal range query, in time $O(\log^{3} n + h)$. Here $h$ is the size of the output.
This work improves the result of (Brass et al. 2013) \cite{brass} that
builds a data structure that uses $O(n \log^{2} n)$ space and has a 
$O(\log^{5} n + h)$ query time. 
Additionally, we show that our data structure can be modified slightly to
solve other related problems. For instance, for counting the number of
points on the convex hull in an orthogonal query rectangle, we propose an
$O(n \log^{2}n)$ space data structure that can be queried upon in $O(\log^{3} n)$ time. 
%This work proposes the first data structure for counting the number of
%points in a query range. -- NOT SURE OF THIS STRONG CLIAM -- KK
We also propose a $O(n \log^{2} n) $ space data structure that can compute the $area$
and $perimeter$ of the convex hull inside an orthogonal range query in $O(\log^{3} n$) time.

\keywords{ Convex Hull, Plane, Orthogonal range, Reporting, Counting, Area, Prerimeter }
\end{abstract}

\section{Introduction}

\noindent Range searching is one of the most commonly studied topics in
spatial databases and computational geometry.  Informally, range searching can
be stated as follows: Given a point set $P$ we wish to pre-process $P$ into a
data structure such that given any rectangular query region $q$, we can
efficiently report the points in $P \cap q$ or count their number in $P \cap
q$. Range aggregate geometric querying is a variant of range searching where
we calculate a geometrical function $f$ inside $P\cap q$. In this work we
efficiently compute $f(P\cap q)$ where the function $f$ is to compute the
convex hull for the point set $P\cap q$. Range aggregate querying has been
studied recently by Kalavagatttu et al.\cite{Anil}, Das et al. \cite{asdas}
and Brodal et al. \cite{brodal} for maximal points.
%It has wide applications in geographic information systems, CAD tools, database retrieval, VLSI etc. 

However, advances in the creation and archival of digital information have led
to an information explosion and therefore the number of objects inside a query
range can be huge. In such cases, reporting a sample of the result set is
preferred. In this paper, we use the concept of convex hull query for
reporting the boundary points. Reporting of convex points can be useful in
situations where we need a shape with minimum area/perimeter for a set of
points. Therefore, it may have applications in spatial databases \cite{Tao}, computer graphics and computer
vision.

In this work, we propose a data structure to solve the problem of reporting
convex hull points in an orthogonal query rectangle.  We also study the
problem of counting the number of points on the convex hull inside $P \cap q$,
and computing its area/perimeter.
\subsection{Problem Definitions}
\label{def}
In this section, we formally define the problems and related terminology.
Given a set of $n$ points $P = \{ p_{1},\ldots,p_{n} \} $ in
$\mathbb{R}^{2}$. We use $x(p)$ to represent the $x$ co-ordinate of a point
$p$ and $y(p)$ to represent its $y$ co-ordinate.
The convex hull of any point set $S$ 
in the Euclidean plane is the smallest convex polygon that contains $S$.
Henceforth, when we refer to convex hull points,
we essentially mean the points on the boundary of the convex hull.
We are given a set $P$ of $n$ points in $\mathbb{R}^{2}$ and a query $q = [x_{lt},x_{rt}]\times{}[y_b, y_t]$.
We find the convex hull on the point set $P \cap q$, denoted by $ch(P \cap q)$. We also denote points with minimum $x$
co-ordinate, maximum $x$ co-ordinate, minimum $y$ co-ordinate and maximum $y$ co-ordinate of point set $P \cap q$
by $x_{min}$, $x_{max}$, $y_{min}$ and $y_{max}$ respectively. 

In this work, we study the following problems:
 We wish to pre-process $P$
into a data structure such that given an orthogonal query region $q$, we can efficiently

{\em Problem 1:} {\bf report} the points on convex hull of $P \cap q$.

{\em Problem 2:} {\bf count} the number of points on the convex hull of $P \cap q$.

{\em Problem 3:} find the {\bf area} of the convex hull of $P \cap q$.

{\em Problem 4:} find the {\bf perimeter} of the convex hull of $P \cap q$.
\subsection{Previous Work} \label{prev-work} The convex hull for a static data
set of two-dimensional points can be computed in $O(n \log h)$ time
\cite{opchan} where $h$ is the number of points on the convex boundary.
Gupta et al. discuss non-intersecting queries on aggregated geometric data \cite{gupta}.
Brass et al. gave the first solution for {\em Problem 1} that takes $O(n \log^{2} n)
$ space, $O(n \log^{3} n) $ preprocessing time and $O(\log^{5} n + h ) $ query
time \cite{brass}.  For any given orthogonal range query on a standard
2d-range tree they identify $O( \log^{2} n)$ disjoint canonical convex
hulls. There are $O( \log^{4} n)$ pairs of disjoint convex hulls and they
compute the tangent for each pair of disjoint convex hulls.  Computing
tangents between each pair of local convex hulls takes $O( \log n)$ time \cite{kirkpatrick}. They
use a method similar to the gift-wrapping algorithm.  Therefore, total query
time is $O( \log^{5} n + h)$.
\subsection{Our Results} \label{our-results}
The following table summarizes the results presented in this paper.
\begin{center}
    \begin{tabular}{ | l | l | l | l | p{2cm} |}
    \hline
    \textbf{Query Type} & \textbf{Query Time} & \textbf{Preprocessing Time} &  \textbf{Storage Space} & \textbf{Theorems} \\ \hline
    Reporting & $O(\log^{3} n + h)$ & $O(n \log^{3} n)$ & $O(n \log^{2} n)$ & Theorem 1 \\ \hline
    Counting & $O(\log^{3} n)$ & $O(n \log^{3} n)$ & $O(n \log^{2} n)$ & Theorem 2 \\ \hline
    Area & $O(\log^{3} n)$ & $O(n \log^{3} n)$ & $O(n \log^{2} n)$ & Theorem 3 \\ \hline
    Perimeter & $O(\log^{3} n)$ & $O(n \log^{3} n)$ & $O(n \log^{2} n)$ & Theorem 4 \\ \hline
    \hline
    \end{tabular}
\end{center}
The rest of the paper is organized as follows. In Section \ref{Basic idea}, we
give the details of the preprocessing and the query algorithm for reporting
convex points inside an orthogonal range query. In Section \ref{other} we
study the problem of counting convex hull points inside $P \cap q$ and the
problem of computing the area/perimeter of the convex hull inside $P \cap
q$. We discuss future work and conclude in Section \ref{conclusion}.
\section{Our Solution}\label{Basic idea}
In this section we explain our solution to the problem of reporting the convex hull points inside $P \cap q$.
In Section \ref{preprocessing} we describe how to construct the required data structure of 
size $O(n \log^{2} n)$ in time $O(n \log^{3} n)$ on a static point set $P$.
In Section \ref{query-al} we explain how to report points on the convex hull inside $P \cap q$ for a monotonic chain from $x_{max}$ to $y_{max}$.
\begin{figure}
\vspace{-0.5cm}
\centering
\includegraphics[scale=1.0]{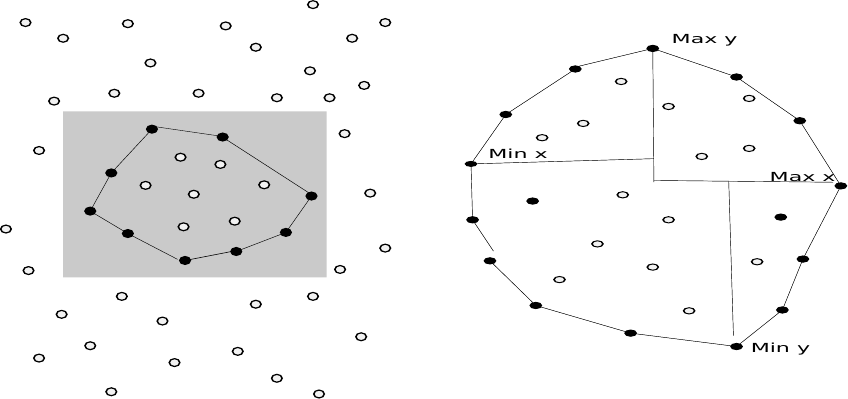}
\caption{($a$) Convex hull inside an orthogonal range query (the shaded region is
  the query). ($b$) Convex hull with four monotone chains}
%($b$.)Problem of finding Convex hull is divided into four subproblems
\label{fig1}
\vspace{-0.5cm}
\end{figure}
Points on the convex hull can be broken into four monotone chains, namely
maximal chain from $x_{max}$ to $y_{max}$, maxY-minX chain from $y_{max}$
to $x_{min}$, minimal chain from $x_{min}$ to $y_{min}$ and
minY-maxX chain from $y_{min}$ to $x_{max}$ as shown in Figure \ref{fig1}($b$).
Area of such a convex hull gets divided into four quadrants Q1, Q2, Q3 and Q4 as shown in Figure \ref{fig1}.
In this work we present an algorithm to construct the maximal chain from
$x_{max}$ to $y_{max}$.  A similar approach can be applied for the other
monotone chains.

\subsection{Preprocessing}\label{preprocessing}
\begin{figure}
\vspace{-0.7cm}
\centering
\includegraphics[scale=0.30]{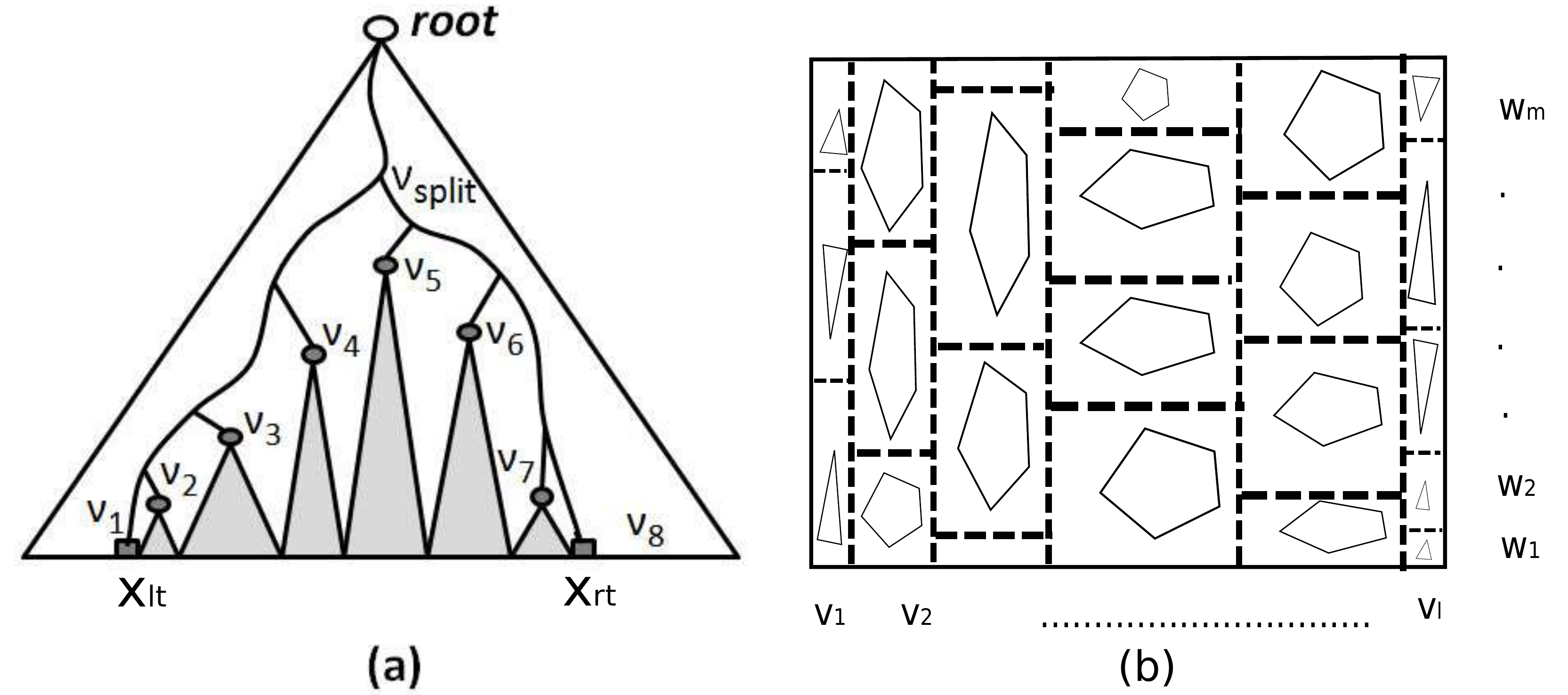}
\caption{($a$.)Range tree with the canonical nodes highlighted ($b$.) Processing an orthogonal range query}
\label{fig3}
\vspace{-0.7cm}
\end{figure}

Our solution uses a $2d$-range tree as described in \cite{berg}.  Each
internal node of the primary tree $T_{x}$ represents a horizontal range
$[x_{i}, x_{j}]$ for $i \neq j$.  The set of points rooted at an internal node
$v$ are represented by $S(v)$. For each internal node $v$ of $T_{x}$ we have a
secondary(binary) tree $T_{y}(v)$ such that leaves of tree $T_{y}(v)$ store
the points in $S(v)$ in non-decreasing order of their $y$-coordinate.  Each
internal node of the secondary tree $T_{y}(v)$ represents a vertical range
$[y_{i}, y_{j}]$ for $i \neq j$.  Given a query $q = [x_{lt}, x_{rt}] \times
[y_{b}, y_{t}]$, we search $x_{lt}$ and $x_{rt}$ in the primary tree to get canonical nodes
% two paths $\pi_{1}$ and $\pi_{2}$ from root to leaves respectively. Let
% $v_{split}$ be the least common ancestor of $\pi_{1}$ and $\pi_{2}$. Let 
$ v_{1}, v_{2}, \ldots, v_{i},\ldots, v_{l} $.
% be the roots of right subtrees and left subtrees of $\pi_{1}$ (from $x_{lt}$ to
% $v_{split}$) and $\pi_{2}$ (from $v_{split}$ to $x_{rt}$) respectively 
as shown in Figure \ref{fig3}. Here $l=O(\log n)$ is the height of the primary
tree. Therefore, the range query $[x_{lt},x_{rt}]$ is divided into $l=O(\log
n)$ canonical subsets $ S(v_{1}), S(v_{2}),\ldots, S(v_{l})$.
Similarly we search for $y_{b}$ and $y_{t}$ in the secondary tree
$T_{y}(v_{i})$ for $i=1,2,3,\ldots,l$ to get canonical nodes
% two paths $\sigma(1,i)$ and $\sigma(2,i)$ from root to leaves respectively. Let $w(i,split)$ be the least
% common ancestor of $\sigma(1,i)$ and $\sigma{(i,2)}$.  Let 
$w(i,1),w(i,2),\ldots, w(i,j),\ldots, w(i,m)$
% be the roots of the right and left subtrees of $\sigma(1,i)$(from $y_{b}$ to
% $w(i,split)$ and $\sigma{(i,2)}$ (from $w(i,split)$ to $y_{t}$) respectively.
Here $m=O(\log |S(v_{i})|)$ is the height of the secondary
tree. Therefore, in every secondary tree, the range query $[y_{b},y_{t}]$ gets
divided into $m=O(\log n)$ canonical subsets $S(w(i,1))$, $S(w(i,2))$,$\ldots$, $S(w(i,m))$.
% For each  internal node $w(i,j)$ of each secondary tree we store a convex hull $CH(S(w(i,j))$.
For each internal node $w$ of the secondary tree, we compute the convex hull over the set $S(w)$.
We store points of this convex hull in an array $ch_{w}$
in the anti-clockwise direction starting from the point with the maximum $y$-coordinate. 
We also find the point $pmax_{w(i,j)}$ with maximum $x$ co-ordinate from the set $S(w(i,j))$ and store it at each internal node $w$.
Therefore, any given query $q = [x_{lt}, x_{rt}] \times [y_{b}, y_{t}]$ gets divided into $O(\log^{2} n)$ disjoint canonical subsets and 
for each of these subsets we have stored a convex hull $ch_{w(i,j)}$ on the set $S(w(i,j))$.

It takes $O(|S(w)|\log |S(w)|)$ time to compute convex hull $ch_{w}$ over set
$S(w)$ and $O(|S(w)|)$ space to store $ch_{w}$ at internal node $w$ of
secondary tree $T_{y}(v)$. Therefore it takes
$\displaystyle\sum\limits_{\forall w \in T_{y}(v)}O(|S(w)| \log |S(w)|) =
O(|T_{y}| \log^{2}|T_{y}|)$ preprocessing time and
$\displaystyle\sum\limits_{\forall w \in T_{y}(v)}O(|S(w)|) = O(|T_{y}|
\log|T_{y}|)$ space for secondary tree $T_{y}(v)$. For any given query
$[x_{lt}, x_{rt}]$ on the primary tree $T_{x}$ we have $l=O(\log n)$ secondary
trees. Therefore it takes $\displaystyle\sum\limits_{\forall v \in
  T_{x}}O(|T_{y}(v)| \log^{2}|T_{y}(v)|) = O(n \log^{3 }n)$ preprocessing time
and $\displaystyle\sum\limits_{\forall v \in T_{x}}O(|T_{y}(v)|
\log|T_{y}(v)|) = O(n \log^{2} n)$ storage space.

\begin{lemma}\label{lem1}
  Given a set $P$ of $n$ points in $\mathbb{R}^{2}$, we can pre-process $P$
  into a data structure that takes $O(n \log^{2} n)$ storage space and $O(n
  \log^{3} n)$ pre-processing time (as explained above).
\end{lemma}

\subsection{Query Algorithm}\label{query-al}%to store all $O(\log^{2} n)$ convexhulls in list $L1$
In this algorithm we use two stacks, a hull stack $HS$ and a tangent stack $TS$. An element of hull stack
$HS$ is a pointer to some canonical convex hull $ch(w(i,j))$. An element of tangent stack $TS$ is a tuple
$t=(i_{1},i_{2})$ where $i_{1}$ and $i_{2}$ are indices of points on two different convex hulls $C_{1}$ and
$C_{2}$ as shown in Figure \ref{fig4}.
Given an orthogonal range query $q = [x_{lt}, x_{rt}] \times [y_{b}, y_{t}]$, the query algorithm
for reporting convex hull points in $P \cap q$ is as follows:
\begin{enumerate}
\item 
Express the range of $x$ co-ordinates in $[x_{lt}, x_{rt}]$ as the
disjoint union of $l$ = $O(\log n)$ canonical subsets. Let the canonical
nodes be $ v_{1}, v_{2},\ldots, v_{l} $ from left to right in that order, as
shown in Figure \ref{fig3}($a$).

\item\label{intial-step} Consider a node $v_{l}$ of the primary tree
  $T_{x}$. Make a range query $[y_{b}, y_{t}]$ on the associated secondary
  tree $T_{y}(v_{l})$ to find canonical nodes $ w(i,1), w(i,2)$, $\ldots$,
  $w(i,m)$. Do a linear search on $x(pmax_{w(i,1)})$, $x(pmax_{w(i,2)})$,
  $\ldots$, $x(pmax_{w(i,m)})$ to find point $X_{maxw}$ with maximum $x$
  co-ordinate. Then traverse the canonical nodes from right to left, starting
  from $v_{l}$ back to $v_{1}$, as shown in Figure \ref{fig3}($a$).
% Find the point $pmax$ with maximum $x$ co-ordinate using binary search on array $X_{v_{l}}$[\ ].
  Initialize $i\leftarrow l$ and $y_{low} \leftarrow y(X_{maxw})$.

\item\label{repeat-step}
Consider a node $v_{i}$ of the primary tree
  $T_{x}$. For each internal node $v_{i}$ of $T_{x}$ there is an associated
  secondary tree $T_{y}(v_{i})$. Make a range query $[y_{low}, y_{t}]$ on
  secondary tree $T_{y}(v_{i})$ to find canonical nodes $ w(i,1),
  w(i,2),\ldots, w(i,m)$, as shown in the Figure \ref{fig3}($b$) where
  $m = O(\log |T_{y}(v_{i})|)$.

\item\label{secondtree}
Consider the array $ch_{w(i,1)}$. %which is the point with maximum $y$-coordinate.
If  the array $ch_{w(i,1)}$ is empty then go to step \ref{primarytree},
otherwise set $y_{low} \leftarrow y(ch_{w(i,m)}[1])$ and then traverse the canonical nodes from bottom to top,
starting from $w(i,1)$ back to $w(i,m)$ as shown in Figure \ref{fig3}($b$) as follows:\\

{\em for} $j\leftarrow 1$ to $m$ call Algorithm Merge($ch_{w(i,j)}$,$HS$,$TS$) (see Section 2.2.1).\\

\item\label{primarytree}
At this point, we have processed the nodes {$ v_{l}, v_{l-1},\ldots,v_{i} $}. 
% If $ytop_{i}$ exists, then set $y_{low} \leftarrow y(ytop_{i})$. 
Set $i \leftarrow i-1$ and if $i \geq 1$, move to the node $v_{i}$ and
goto step \ref{repeat-step}, else exit.

\item
Call Algorithm Report($HS$,$TS$)(see Section 2.2.2) to report the points on the convex monotone from 
maximum $x$ to maximum $y$ inside $P \cap q$.%\ref{reporting}.
\end{enumerate}
\subsubsection{2.2.1  Merge Algorithm}\label{merge-al}
\begin{figure}
\vspace{-0.5cm}
\centering
\includegraphics[scale=0.30]{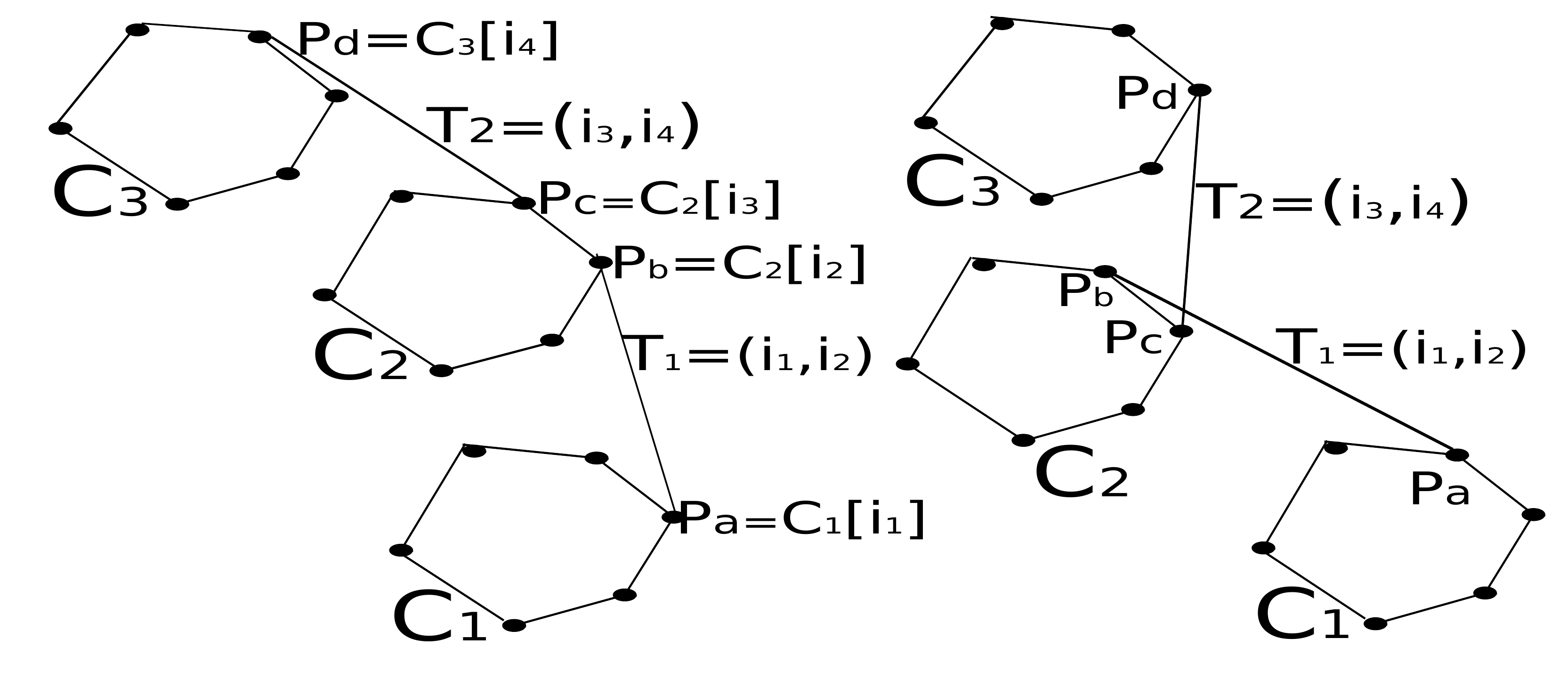}
\caption{($a$)Positive Orientation ($b$) Negative Orientation}
\label{fig4}
\vspace{-0.7cm}
\end{figure}

\paragraph{}
In this section we explain the Algorithm \emph{Merge()} used for merging canonical
convex hulls for any given query.  Note that the algorithm proposed is similar
to the Graham Scan algorithm \cite{graham} where we combine points which are
sorted on $x$ co-ordinate.
Instead of points we have disjoint convex hulls sorted on non-increasing $x$ co-ordinate in the stack $HS$.
\begin{algorithm}\label{merging}
\KwData{$ch_{w(i,j)}$, $HS = \phi$, $TS = \phi$}
\KwResult{Updated Stacks $HS$ and $TS$ }
\While{$size(HS)\geq 2$}{
$C_{1} \leftarrow secondtop(HS)$;$C_{2} \leftarrow top(HS)$;$C_{3} \leftarrow ch_{w(i,j)}$\;
find points $P_{a}$ and $P_{b}$ on tangent $T_{1}$=($i_{1}$,$i_{2}$) joining hulls $C_{1}$ and $C_{2}$\;
find points $P_{c}$ and $P_{d}$ on tangent $T_{2}$=($i_{3}$,$i_{4}$) joining hulls $C_{2}$ and $C_{3}$\;
\eIf{$orient(p_{a},p_{b},p_{d})\leq 0$ }{\tcc{(refer Figure \ref{fig4}($b$))}
 pop($HS$); pop($TS$)\;
}{\tcc{(refer Figure \ref{fig4}($a$))}
push($T_{2}$);
break\;}
}
push $ch_{w(i,j)}$ onto stack $HS$\;
\caption{\em Merge()}
\end{algorithm}
In line 1 of the algorithm, a while loop checks whether hull stack $HS$ has
sufficient elements to continue.  In line 2 we store hulls at $secondtop(HS)$,
$top(HS)$ and $ch(w(i,j))$ in variables $C_{1}$, $C_{2}$ and $C_{3}$.  In line
3 we find points $p_{a}\leftarrow C_{1}[i_{1}]$ and $p_{b}\leftarrow
C_{2}[i_{2}]$ incident with tangent $T_{1}$=($i_{1}$,$i_{2}$) between hulls
$C_{1}$ and $C_{2}$ as shown in the Figure \ref{fig4}.  Computing such a tangent takes $O(\log(n_{1}+n_{2}))$
points where $n_{1}=|C_{1}|$ and $n_{2}=|C_{2}|$ are sizes of the convex
hulls.  Similarly in line 4 we find points $p_{c}\leftarrow C_{2}[i_{3}]$ and
$p_{d}\leftarrow C_{3}[i_{4}]$ incident with tangent $T_{2}$=($i_{3}$,$i_{4}$)
between hulls $C_{2}$ and $C_{3}$ as shown in the Figure \ref{fig4}.  In line 5 we compute the orientation of
the points $p_{a}$, $p_{b}$ and $p_{d}$ using function $orient()$(a primitive opeation in computational Geometry). If the orientation of the points is
clockwise (negative) as shown in Figure \ref{fig4}(b) then we pop out an element from the stack $HS$ and
the stack $TS$ in line 6, else we push tangent $T_{2}$ into stack $TS$ and break out of the loop in line 8.
Every time an element gets pushed onto the stack we iterate the above
procedure on the top two elements of stack $HS$ and $ch_{w}$ to discard canonical set of
convex hulls $ch(w(i,j))$ that do not contribute points to $ch(P \cap q)$.
For every call to Algorithm \emph{Merge()} there
can be zero or more pop() operations on stack $HS$. An element once popped out
of the stack $HS$ never gets pushed in again. Therefore, the total number of
push and pop operations is $O(|HS|)$. For each pop and push operation on
the stack $HS$, the required tangent can be computed in $O(\log n)$ time.
Finding the tangent between any two canonical convex hulls takes $O(\log(n1+n2)) = O(\log n)$
time \cite{kirkpatrick}.  Therefore, the merging algorithm takes no more than
$O(|HS|\cdot \log n)$ time where $|HS|$ is size of the stack. We end this section with the following lemma.

\begin{lemma}\label{lem2}
Given the stack $HS$, the stack $TS$ and the hull $ch_{w(i,j)}$ as input to {\bf Algorithm Merge()}, it takes 
$O(|HS|\cdot \log n)$ time where $|HS|$ is size of the stack $HS$.
\end{lemma}

\subsubsection{2.2.2  Reporting algorithm}\label{putting}
In this section we explain the algorithm \emph{Report()} used for reporting
points on the convex hull.  This reporting algorithm is the last step of the
query algorithm.
\begin{algorithm}\label{reporting}
\KwData{Stack $HS$,Stack $TS$}
\KwResult{Convex hull points in $P \cap q$ }
$ch \leftarrow pop(HS$);
index $i \leftarrow size(ch)$\;
Report the point $ch[i]$ with maximum $y$ co-ordinate\;
\While{$size(HS)>0$}{
tangent $t \leftarrow pop(TS$)\;
index $j \leftarrow t(i_{2}$)\;
Report points from $ch[i]$ to $ch[j]$\;
$i \leftarrow t(i_{1}$)\;
array $ch \leftarrow pop(HS)$; 
}
Report points from  ch[$i$] till $x_{max}$\ in array $ch$\;
\caption{Report()}
\end{algorithm}
\vspace{-0.7cm}
We report points from canonical convex hulls stored in
stack $HS$ using information about tangents stored in the stack $TS$.
In line $1$ of the algorithm we pop out a hull from the stack $HS$ and index $i$
is assigned with the size of array $ch$. In line $2$ we report the point $ch[i]$ with maximum $y$ co-ordinate.
In line $3$ we check the while loop condition, if the stack is not empty then we enter the while loop.  In line 4
we pop out tangent $t$ from the stack $TS$ and in line 5 we assign $j$ with
the first index $i_{2}$ of the tangent.  In line 6 we report points from
ch[$i$] to ch[$j$].  In line 7 we assign $i$ with the second index $i_{1}$ of
tangent $t$ and in line 8 we pop out a canonical convex hull from the stack $TS$.

In the algorithm \emph{Report()} we iterate till the stack $HS$ is empty and in each iteration of the while loop
we report a set of points. Let total number points reported by Algorithm \emph{Report()} be $h$.
It takes $O(|HS|)$ time for the algorithm \emph{Report()} because  while loop gets iterated $|HS|$ times.
Therefore, it takes  $O(|HS| + h_{1} )$ time for algorithm \emph{Report()}.
We end this section with the following lemma.
\begin{lemma}\label{lem3}
Given the stack $HS$ and the stack $TS$ as input to {\bf Algorithm Report()}, it takes 
$O(|HS| + h_{1} )$ time where $|HS|$ is size of the stack $HS$ and $h$ is the number of points
reported.
\end{lemma}

\subsection{Putting Everything Together}
We will now prove the following theorem:
\begin{theorem}
  Given a set $P$ of $n$ points in $\mathbb{R}^{2}$, we can pre-process $P$
  into a data structure of size $O(n \log^{2} n)$ in time $O(n \log^{3} n)$
  such that, given an orthogonal range query $q = [x_{l}, x_{r}] \times
  [y_{b}, y_{t}]$, we can report the points of the convex hull inside $P \cap
  q$ in time $O(\log^{3} n + h)$, where $h$ is the number of convex hull
  points reported.
\end{theorem}
$\textbf{Proof}$ : Lemma \ref{lem1} for preprocessing indicates the storage space
used by our data structure and the preprocessing time to build it.  Now we analyze the
complexity of our query algorithm explained in Section \ref{query-al}.  In
step 1 it takes $O(\log n)$ time to find $l = O(\log n)$ canonical nodes
because the height of primary tree $T_{x}$ is $O(\log n)$.  In step 2 it takes
$O(\log |S(v_{l})|)$ time to find canonical subset of nodes $ w(l,1), w(l,2)$,
$\ldots$, $w(l,m)$. It takes $m = O(\log |S(v_{l})|)$ time to perform a linear
search on $x(max_{w(i,1)})$, $x(max_{w(i,2)})$, $\ldots$, $x(max_{w(i,m)})$ to
find point $p_{xmax}$ with maximum $x$ co-ordinate.  In step 3 we consider
each $v_{i}$ from $i \leftarrow l$ to 1. In $i^{th}$ iteration of step 3 we
spend $O(\log |S(v_{i})|) = O(\log n)$ time finding the canonical subset of nodes $ w(i,1),
w(i,2)$, $\ldots$, $w(i,m)$ by making an updated query on the associated
secondary tree $T_{y}(v_{i})$. Therefore, the total time spent in finding all
canonical sets of nodes for any given query is $O(\log |S(v_{1})|)+O(\log
|S(v_{2})|)+\ldots+O(\log |S(v_{l})|)$ 
for $i = 1,2,\ldots,l$ then total time is $O(l\cdot \log n) = O(\log^{2}n)$. In step 4 at
most $m = O(\log^{2} n)$ calls are made to Algorithm \emph{Merge()}. Therefore total calls made to algorithm \emph{Merge()} is $m \cdot l = O(\log^{2}n)$.
From Lemma \ref{lem2} we know that time taken for Algorithm \emph{Merge()} is $O(|HS|\cdot \log n)$
Therefore, it takes $O(\log^{3} n)$ time to find updated $HS$ that contains convex hulls that contribute
at least one point to $ch(P\cap q)$.
In step 6 of the query algorithm a call to the algorithm Report($HS$,$TS$) is made.
According to Lemma \ref{lem3} it takes $O(|HS| + h_{1} )$ time where $|HS|$ is size of the stack $HS$ and $h$ is the number of points
reported. Therefore, it takes $O(|HS| + h_{1} )$ time for step 6 to execute where $|HS|=O(\log^{2} n)$. Recall that time taken for step $3-5$ is $O(\log^{3} n)$.
Therefore it takes $O(\log^{3} n)+O(\log^{2} n + h_{1})=O(\log^{3} n + h_{1})$ to report the points of $ch(P \cap q)$ in the first quadrant Q1 (see Figure \ref{fig1}).

\section{Related Problems}\label{other}
In this section we study related problems such as counting the number of
convex hull points and finding the area/perimeter of the convex hull inside $P
\cap q$.  One may argue that it is not necessary to study these problems
separately once we have reported the points of $ch(P \cap q)$ in query time
$O(\log^3 n + h)$ using the algorithm described in Section \ref{query-al}
%The Counting, Area and Perimeter are trivial problems once we had computed convex hull $CH(P \cap q)$ for any given query. Our solution$O(\log^{3} n + h)$ 
where $h$ is the total number of points on the convex hull. However, $h$ can
be as large as $O(n)$ for some queries.
%However, if these problems are solved using points reported from {\em
%Algorithm report()} then it takes $O(\log^{3} n + h)$ time.
If these problems can be solved independent of $O(h)$ then the running time is
unaffected by output size.  With a few clever modifications on the data
structure explained in Section \ref{preprocessing} and the algorithm proposed
in Section \ref{query-al} we can achieve results that are independent of
output size for the above mentioned problems.

\subsection{The Counting Problem}\label{Count}
Most of the preprocessing of the point set $P$ for the counting problem is the
same as the preprocessing described in Section \ref{preprocessing}. In
addition to it, at each internal node $w$ of every secondary tree $T_{y}(.)$
we store three indices $index_{xmax}$, $index_{xmin}$ and $index_{ymin}$ where
$ch_{w}[index_{xmax}]$, $ch_{w}[index_{xmin}]$ and $ch_{w}[index_{ymin}]$ are
points with maximum $x$ co-ordinate, minimum $x$ coordinate and minimum $y$
co-ordinate.  To get the stack $HS$ which contains all the canonical convex
hulls that contribute to convex hull $ch(P\cap q)$ we use same Query Algorithm
of Section \ref{query-al}. Below is the algorithm for counting the number of
points on the convex hull of set $S(P \cap q)$. To count the points in $ch(P \cap q)$, 
algorithm counting() can be used as subroutine in step 6 of the query algorithm.
\begin{algorithm}
 \KwData{stack $HS$,stack $TS$}
 \KwResult{$count$ of the points of $ch(P \cap q)$ from maximum $y$ to maximum $x$ }
    array $ch \leftarrow pop(HS$);
    index $i \leftarrow size(ch)$\;
    $count \leftarrow 1$\;
  \While{$size(HS)>0$}{
    tangent $t \leftarrow pop(TS)$\;
    index $j \leftarrow t(i_{2})$\;
    $count \leftarrow count + (j - i + 1)$\;
    $i \leftarrow t(i_{1})$\;
    array $ch \leftarrow pop(HS$); 
   }
     $count \leftarrow count + (index_{xmax} - i + 1)$\;
 \caption{Counting()}
\end{algorithm}
\vspace{-0.7cm}
Algorithm \emph{Counting()} is similar in spirit to the
Algorithm {\emph{Report()} of Section 2.2.2. In
Step 2 of this algorithm $count$ is initialized to $1$ because we count all the
points from point $ch[j]$ till the point with maximum $y$ co-ordinate in $P \cap q$. In Step
6 the variable $count$ gets updated with the number of points that the current
canonical hull contributes to the output. This is computed as the difference
$i-j+1$ in each iteration of while loop. This process is repeated until the
stack $HS$ is empty.  In step 10 $count$ gets updated with the difference
$index_{xmax} - i + 1$ after exiting from the while loop, where $index_{xmax}$
is the index of point with the maximum $x$ co-ordinate in the array $ch$.

It can be seen that Algorithm \emph{Counting()} runs in $O(|HS|) =
O(\log^{2}n)$ query time to count points of the convex hull $ch(P \cap q)$
from maximum $x$ to maximum $y$. Similar counting algorithms can be used to
find the count of points of the convex hull $ch(P \cap q)$ for the other three
monotone chains (see also Figure \ref{fig1}).
\begin{theorem}
Given a set $P$ of $n$ points in $\mathbb{R}^{2}$, we can pre-process $P$
into a data structure of size $O(n \log^{2} n)$ in time $O(n \log^{3} n)$
such that, given an orthogonal range query $q = [x_{l}, x_{r}] \times
[y_{b}, y_{t}]$, we can count the points of the convex hull inside $P \cap
q$ in time $O(\log^{3} n)$.
\end{theorem}

\subsection{The Perimeter Problem}
For computing the perimeter of the convex hull of the points in $P\cap q$ on a
two-dimensional point set $P$ and an orthogonal range $q$, we perform a
preprocessing phase that is similar to the one described in Section
\ref{preprocessing}. In addition, at every internal node $w$ of every
secondary tree, we store an auxiliary array $P_{w}$ that stores the cumulative perimeter.
$P_{w}[i]=\displaystyle\sum\limits_{1\le j\le i} dist(j,j+1)$ where
$dist(j,j+1)$ is the distance between points $ch_{w}[j]$ and $ch_{w}[j+1]$
respectively.  Below Algorithm  {\em Perimeter()} used for finding the perimeter of the convex
hull $ch(P \cap q)$ from maximum $x$ to maximum $y$.
\vspace{-0.7cm}
\begin{algorithm}
 \KwData{$HS$,$TS$}
 \KwResult{$perimeter$ of $ch(P \cap q)$ from maximum $y$ to maximum $x$}
  convex hull $ch \leftarrow pop(HS$);
  index $i \leftarrow 1$\;
   $perimeter=0$\;
 \While{$size(HS)>0$}{
   tangent $t \leftarrow pop(TS)$\;
   index $j \leftarrow t(i_{2})$\;
   point $a=ch[j]$\;
   $perimeter \leftarrow perimeter+(P_{w}[i]-P_{w}[j])$\;
   $i \leftarrow t(i_{1})$\;
   $ch=pop(HS)$\; 
   point $b=ch[i]$\;
   $perimeter = perimeter + dist(a,b)$\;
 }
  $perimeter \leftarrow perimeter+(P_{w}[i]-P_{w}[index_{xmax}])$\;
  \caption{Perimeter()}
\end{algorithm}
\vspace{-0.7cm}
Algorithm {\em Perimeter()} is similar in spirit to the reporting
algorithm (Algorithm Report) of Section \ref{Count}.
%Above mentioned perimeter algorithm is exactly similar to reporting algorithm of Section 2.2.2, except step 2, step 6, step 7, step 10,step 11 and step 13.
In Step 2 of this algorithm, the variable $perimeter$ is initialized to
$0$. In Step 6 we store point $ch[i_{1}]$ of some hull $C_{1}$ into point $a$.  In
Step 7 the variable $perimeter$ is updated with difference $P_{w}[i]-P_{w}[j]$
in each iteration of while loop. This process is repeated until the stack $HS$
is empty.
%In Step 10 we store point $ch[i_{2}]$ of next hull $C_{2}$ into $b$.
%In Step $perimeter$ gets updated with distance between points $a$ and $b$.
In Step 13 the variable $perimeter$ is updated with difference
$P_{w}[i]-P_{w}[index_{xmax}]$ after exiting from while loop where
$index_{xmax}$ is index of point with the maximum $x$ co-ordinate in array
$ch$. All other steps are similar to counting algorithm in section \ref{Count}

It can be noticed that the Algorithm Perimeter runs in $O(|HS|) =
O(\log^{2}n)$ time to find perimeter of the convex hull $ch(P \cap q)$ from
maximum $x$ to maximum $y$. A similar algorithm can be used to find the
perimeter of the convex hull $ch(P \cap q)$ for other three monotone
chains. (See also Figure \ref{fig1}.)

\begin{theorem}
  Given a set $P$ of $n$ points in $\mathbb{R}^{2}$, we can pre-process $P$
  into a data structure of size $O(n \log^{2} n)$ in time $O(n \log^{3} n)$
  such that, given an orthogonal range query $q = [x_{l}, x_{r}] \times
  [y_{b}, y_{t}]$, we can compute the perimeter of the convex hull inside $P
  \cap q$ in time $O(\log^{3} n)$.
\end{theorem}

\subsection{The Area Problem}
During the preprocessing phase, an auxiliary array $A_{w}$ is stored at each
internal node of every secondary tree.  Each element of such an array stores
the cumulative area $A_{w}[i]=\displaystyle\sum\limits_{3\le j\le i}
aot(1,j-1,j)$ where $aot(1,j-1,j)$ is the area of the triangle between points
$ch_{w}[1]$, $ch_{w}[j-1]$ and $ch_{w}[j]$ respectively.  $A_{w}[1]=0$ because
it represents the area of point $ch[1]$ and $A_{w}[2]=0$ because it is the area
of the line joining points $ch[1]$ and $ch[2]$. Below is an algorithm for computing the area of the convex hull $(P \cap q)$ .
\begin{algorithm}
 \KwData{Stacks $HS$,$TS$,$AR = \phi$}
 \KwResult{Area of the convex hull of Q1 in Figure \ref{fig1}}
  $area_{Q1}=0$\;
  point $fixedPoint$
  array $ch \leftarrow pop(HS$);
  index $i \leftarrow 1$\;
  $push(ch[i])$ onto $AR$\;
  $y(fixedPoint) = y(ch[1])$\;
 \While{$size(HS)>0$}{
   tangent $t \leftarrow pop(TS$)\;
   index $j \leftarrow t(i_{2}$)\;
   $push(ch[j])$ onto $AR$\;
   $area_{Q1} \leftarrow area_{Q1} +(A_{w}[i]-A_{w}[j]) - aot(1,i,j)$\;
   $i \leftarrow t(i_{1}$)\;
   $push(ch[i])$ onto $AR$\;
    ch=pop($HS$); 
 }
  $j \leftarrow index_{xmax}$\;
  $push(ch[j])$ onto $AR$\;
  $area_{Q1} \leftarrow area_{Q1}+(A_{w}[i]-A_{w}[j]) - aot(1,i,j)$\;
  $x(fixedPoint) = x(ch[j])$\;
  point $k1 \leftarrow pop(AR)$\;
  \While{$size(AR) > 0$}{
  point $ k2 \leftarrow pop(AR)$\;
  $area_{Q1} \leftarrow area_{Q1}+aot(fixedPoint,k1,k2)$\;
  $k1  \leftarrow k2$\;
  }
\caption{Area()}% for computing area of the convex hull inside $P \cap q$}
\end{algorithm}
\vspace{-0.6cm}
The  algorithm {\em Area()} finds the area of quadrant Q1 (Figure
\ref{fig1}). Similarly we can find areas of the other quadrants Q2,Q3 and
Q4. The area of $ch(P \cap q)$ can be obtained using the following formula.\\
Area(ch($P \cap q$) = $area_{Q1}+area_{Q2}+area_{Q3}+area_{Q4} - (|x(y_{max})
- x(y_{min})|*|y(x_{max}) - y(x_{min})|)$ where $x_{min}$, $x_{max}$, $y_{min}$ and $y_{max}$
are the points with minimum $x$, maximum $x$, minimum $y$ and maximum $y$ in $P \cap q$.
\begin{theorem}
  Given a set $P$ of $n$ points in $\mathbb{R}^{2}$, we can pre-process $P$
  into a data structure of size $O(n \log^{2} n)$ in time $O(n \log^{3} n)$
  such that, given an orthogonal range query $q = [x_{l}, x_{r}] \times
  [y_{b}, y_{t}]$, we can compute the area of the convex hull inside $P \cap
  q$ in time $O(\log^{3} n)$.
\end{theorem}

\section{Conclusion}\label{conclusion}
In this work, we studied the problem of reporting convex hull points for any
given orthogonal range query, in the \emph{Pointer Machine Model}. We also
solved the problem of counting, area and perimeter in $O(\log^{3} n)$ time.
We restricted the point set to static two-dimensional points. It will be
interesting to see these problems in higher dimensions and dynamic versions of
the problems.  It will be also interesting to study these problems related to convex
hull in the \emph{Multi-shot Model}. It will be interesting to study these problems
in {word-RAM} model and \emph{Cell-probe Model}.

\end{document}